\documentclass[aps,prb,twocolumn,floats,floatfix,showpacs,footinbib]{revtex4-1}

\usepackage{calc}
\usepackage{graphicx}
\usepackage{color}
\usepackage{amsmath, amssymb}
\usepackage[percent]{overpic}
\usepackage{array}
\usepackage{hyperref}

\begin{document}

% *******************************************************************
% *** TITLE PAGE                                                  ***
% *******************************************************************
\title{Anderson localization at large disorder}
\author{L{\'a}szl{\'o} \surname{Ujfalusi}}
\author{Imre \surname{Varga}}
\email[Contact: ]{varga@phy.bme.hu}
\affiliation{Elm{\'e}leti Fizika Tansz{\'e}k, Fizikai Int{\'e}zet, 
Budapesti M{\H{u}}szaki {\'e}s Gazdas{\'a}gtudom{\'a}nyi Egyetem, 
H-1521 Budapest, Hungary}
\date{\today}
\begin{abstract}
The localization of one-electron states in the large (but finite)
disorder limit is investigated. The inverse participation number 
shows a non--monotonic behavior as a function of energy owing to 
anomalous behavior of few-site localization. The two-site 
approximation is solved analytically and shown to capture the 
essential features found in numerical simulations on one-, two- 
and three-dimensional systems. Further improvement has been 
obtained by solving a three-site model. 
\end{abstract}
\maketitle

{\it Introduction --}
The problem of disordered systems has been in the front line of research of condensed matter physics for several decades starting 
from the seminal paper of Anderson~\cite{And58}. Electron localization is an essential phenomenon playing an important 
role especially for low-dimensional systems~\cite{MirlinEvers}. It is commonly known that the spatial extension of the electronic 
states is a monotonous function of disorder and position in the energy-band, however, recent studies~\cite{Johri-Bhatt}
presented new results that call for further understanding.

In this paper we focus on the large disorder limit of the Anderson problem~\cite{And58}, therefore our Hamiltonian reads as
\begin{equation} 
\mathcal{H}=\sum_i \varepsilon_i a_i^\dagger a_i -t \sum_{\left<i,j\right>} \left( a_i^\dagger a_j + a_j^\dagger a_i \right)\quad ,
\label{hamilton}
\end{equation}
where the on-site potential, $\varepsilon_i$ are uniformly distributed over $\left[-\frac{W}{2},\frac{W}{2}\right]$, hence the parameter $W$
characterizes the strength of disorder. We measure the energy in units of the hopping, $t$, which is equivalent of setting $t=1$. 
The problem described by this Hamiltonian has been studied extensively in the past \cite{And58, MirlinEvers}. 
As it is known, there is a critical disorder, $W_c$, beyond which, $W>W_c$, every eigenstate is exponentially localized. In one and two dimensions 
$W_c=0$, meanwhile in three dimensions for the above model $W_c\approx 16.5$ for the states in the band center. 

In the present work we investigate the properties of the states close to the band-edge in the strongly localized regime, $W\gg W_c$. 
The spatial extent of the eigenstates is commonly characterized by the parameter called the inverse participation ratio 
(IPR)~\cite{MirlinEvers,VargaPH}, 
\begin{equation}
I=\sum_{i=1}^{N}|\Psi_i|^4\quad .
\label{ipr}
\end{equation}
For a state extending homogeneously over $k$ sites $I=k^{-1}$, thus $1/I$ tells us the effective number of sites a state extends to,
hence the name. A state localized on one single site would give $I=1$, but extending over the whole system of size $N$, $I=1/N$. Hence any 
states will have an IPR value between these two cases, $1/N<I<1$. In summary the IPR, $I$ is a measure of localization, its inverse a 
measure of extension.

\begin{figure}[tbh]
	\begin {center}
	\begin{tabular}{ c c c}
	\begin{overpic}[height=4cm]{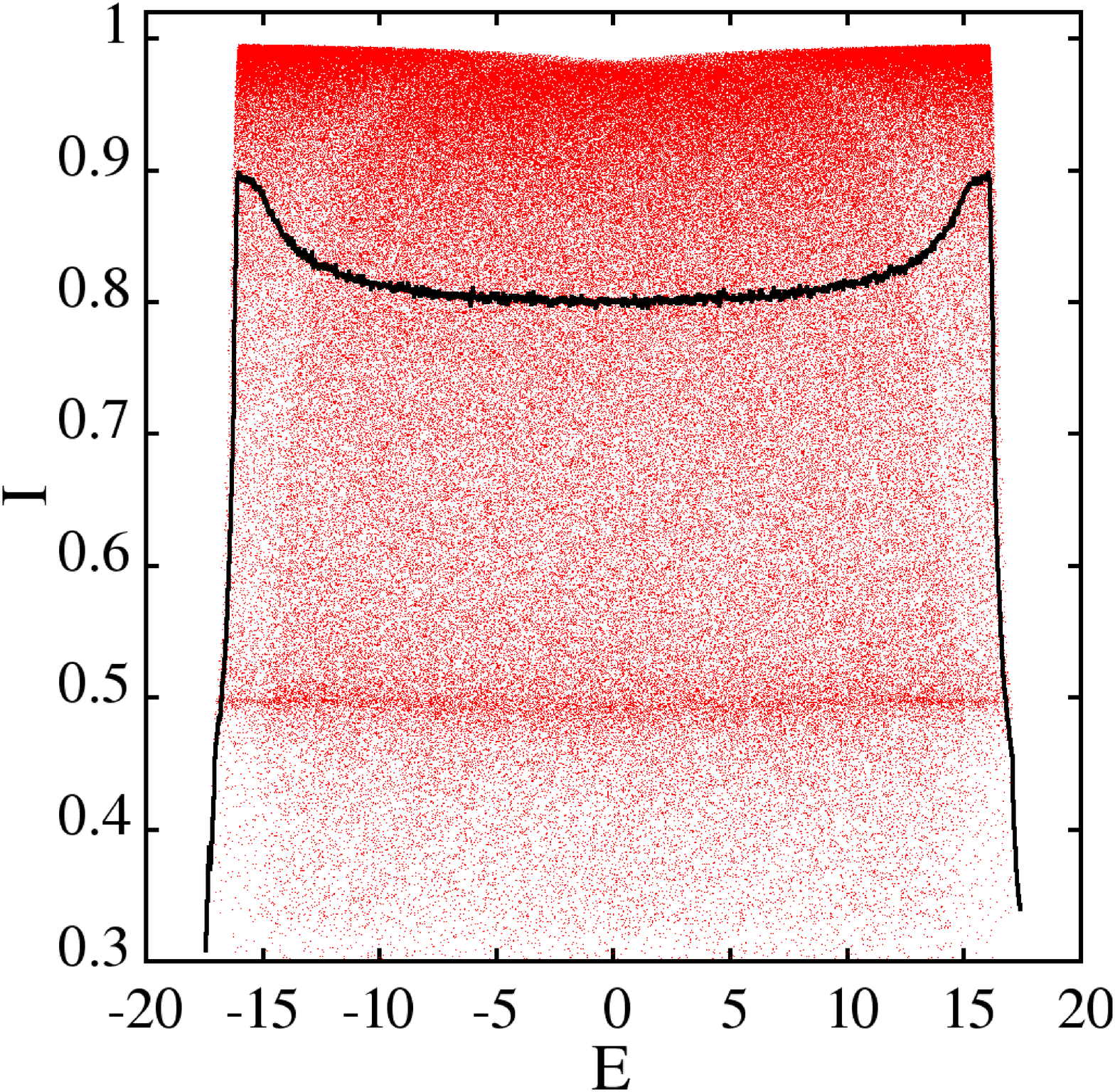} \put(-6,100){(a)} \end{overpic} 
  & \begin{overpic}[height=4cm]{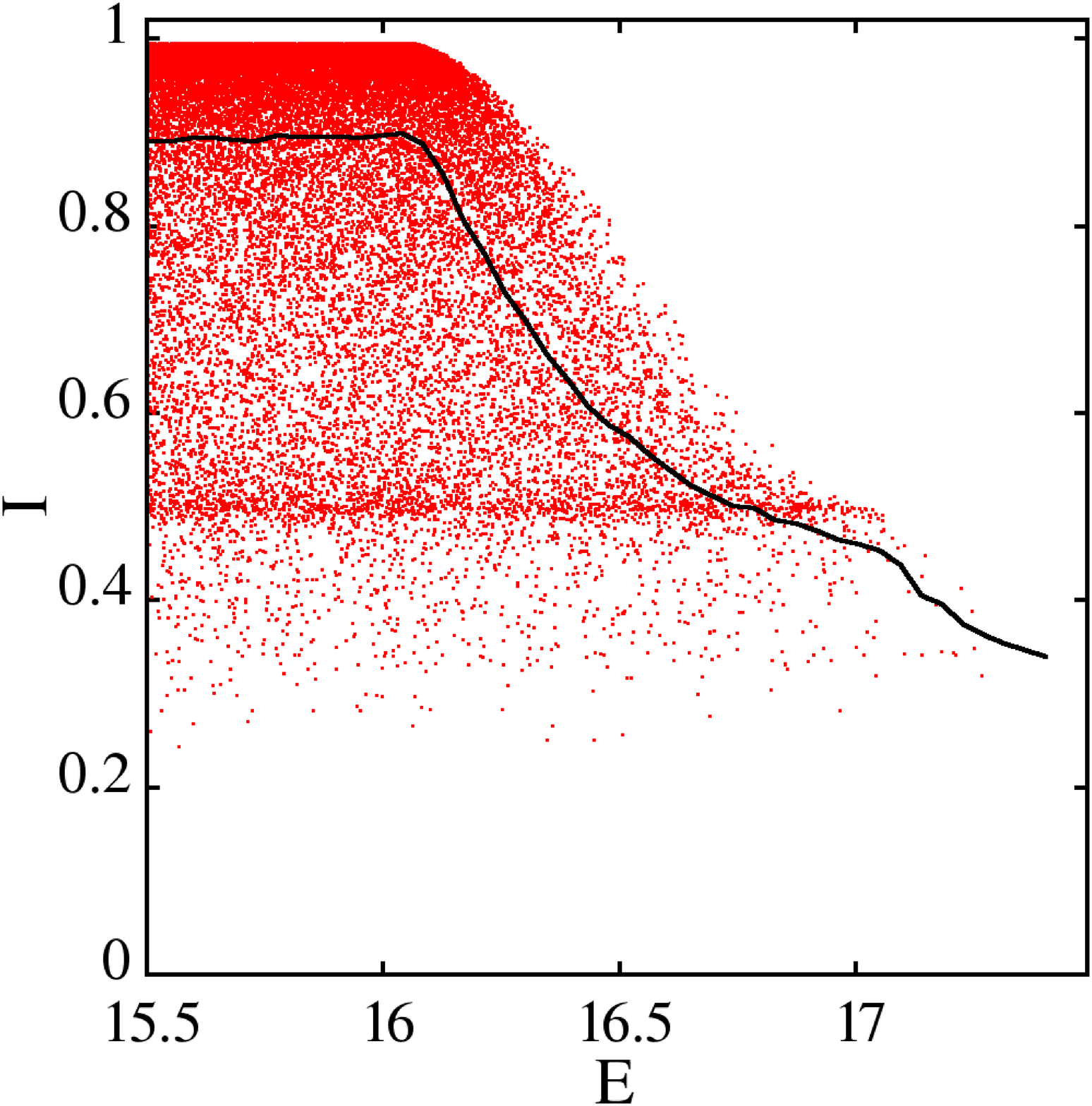} \put(-6,100){(b)} \end{overpic}\\
	\end{tabular}
	\caption{(a) The IPR of the states as a function of energy for $W=32$. The red dots correspond to the states, the black curve is the average. 
             (b) The same as (a) but zoomed to the band edge.}
	\label{fig:1DIPR}
	\end{center}
\end{figure}

For strong disorder the states are expected to extend over a few sites only, therefore we expect IPR values typically of the order of 
$I\to 1$, hence finite size effects will not disturb our numerical simulations and in addition relatively small systems can be used. 
We employed periodic boundary conditions, and linear system size $L=512$ in $d=1$, $L=20$ in $d=2$ and $L=8$ in $d=3$. We computed 
the eigenvalues and the eigenvectors of the Hamiltonian (\ref{hamilton}), and the IPR for every state. We made statistical averaging over 
$M=12500$ realizations, the results of the one dimensional (1D) case are shown in Fig.~\ref{fig:1DIPR} for disorder strength $W=32$. 
In the subsequent part of the paper we present our arguments using the 1D case but extensions for $d=2$ and $d=3$ will be presented, as well. 
In Fig.~\ref{fig:1DIPR}(a) it is shown that the dots are distributed symmetrically around the band center, $E=0$, due to particle-hole symmetry.
In Fig.~\ref{fig:1DIPR} we may also see that the distribution of the IPRs is broad but many of the states seem to have values close to unity, $I=1$, 
and there is a considerable amount of them around $I=1/2$, as well. As a function of energy we can see that moving away from the band-center 
the average IPR, $\langle I\rangle$ first increases in accordance with our expectation of increasing localization towards the band edge. 
However, beyond a certain energy to be discussed later, 
\begin{equation}
E_0 = \sqrt{1+\left(\frac{W}{2}\right)^2}\quad ,
\label{ener}
\end{equation}
the average IPR decreases again, even though the Lyapunov--exponent, the inverse of the characteristic length of the exponential tail of the state,
(not shown here) would further increase undisturbed over this energy scale. This phenomenon has been recently noticed and presented in 
Ref.~\onlinecite{Johri-Bhatt}. This means, that at the band edge the effective size of a state becomes larger, even if the localization 
length decreases. The main aim of the present work is to understand and give an analytical explanation for the behavior of Fig.~\ref{fig:1DIPR}
especially its part (b), showing an empty region of the IPR for large value of the energy together with a decrease of the average IPR, $\langle I\rangle$.

A qualitative explanation of the existence of this region is as follows. If the energy of an eigenstate is bigger than $W/2$ then it must extend at least 
over two sites, because the on-site energies are bounded by $|\varepsilon|\le W/2$, hence for large energies beyond the potential energy the
states should have some additional kinetic energy which can be obtained by allowing their extension over more than one site. 
This is the reason why close to the band edge the states become more extended and hence the IPR has an upper bound. A similar argument 
can be found in~\cite{Johri-Bhatt}.

In order to understand the large-$W$ behavior of the sates first we have to start with the asymptotic behavior, i.e. as $W\to\infty$. In that limit
all the states will be localized to a single site and hopping, i.e. kinetic energy, plays no role. In this case the system is a kind of sum of 
independent sites, therefore from the point of view of a probabilistic description, it is enough to take into account just one site with a random 
on-site energy and one electron on it. The probability distribution for $E$ and IPR of this one-site system is the same as for a large system due
to the independence of sites. The energy is just the random potential energy, thus the model gives us a $W$ wide band, which is very close to reality 
(see Fig.~\ref{fig:1DIPR}), for large enough disorder. Since every state is localized to one site, for every state $I=1$ and that also becomes a 
increasingly better approximation as $W$ increases. But as mentioned above, for finite disorder there is an interesting inner structure in the figures, 
that this one-site model cannot capture. Thus we tried to find a better model. For large enough $W$ the states are strongly localized and extend 
over a few sites, which can be depicted in Fig.~\ref{fig:1DIPR}. For example at $W=32$ most of the IPRs are larger than $1/2$, in other words most 
of the states extend approximately over two sites, so a two- or three-site model should be enough - at least qualitatively - to describe this 
strongly localized system.

\begin{figure}[tbh]
	\begin {center}
	
	\begin{tabular}{ c c }
	\begin{overpic}[height=4cm]{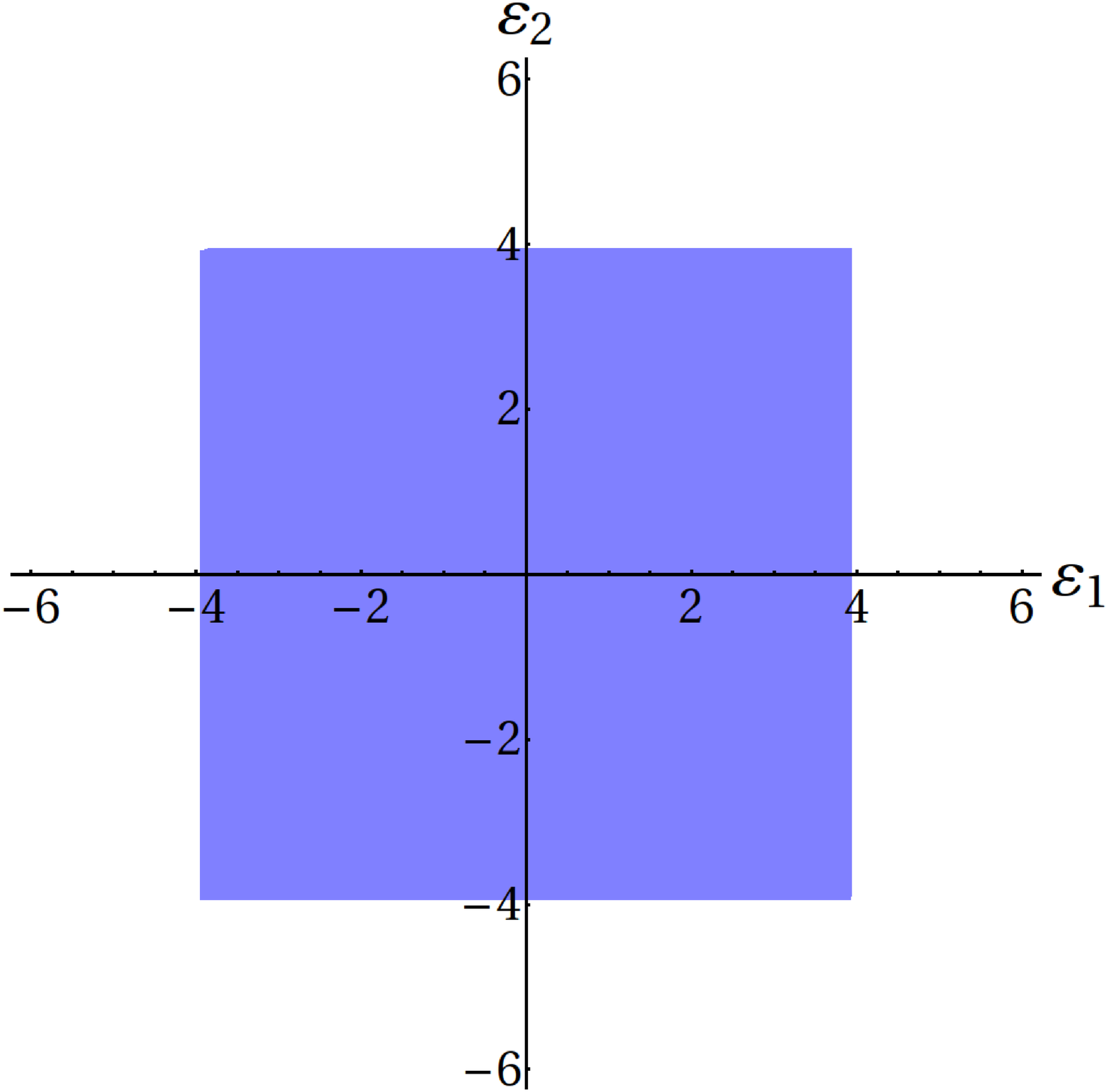} \put(0,90){(a)} \end{overpic} & 
    \begin{overpic}[height=4cm]{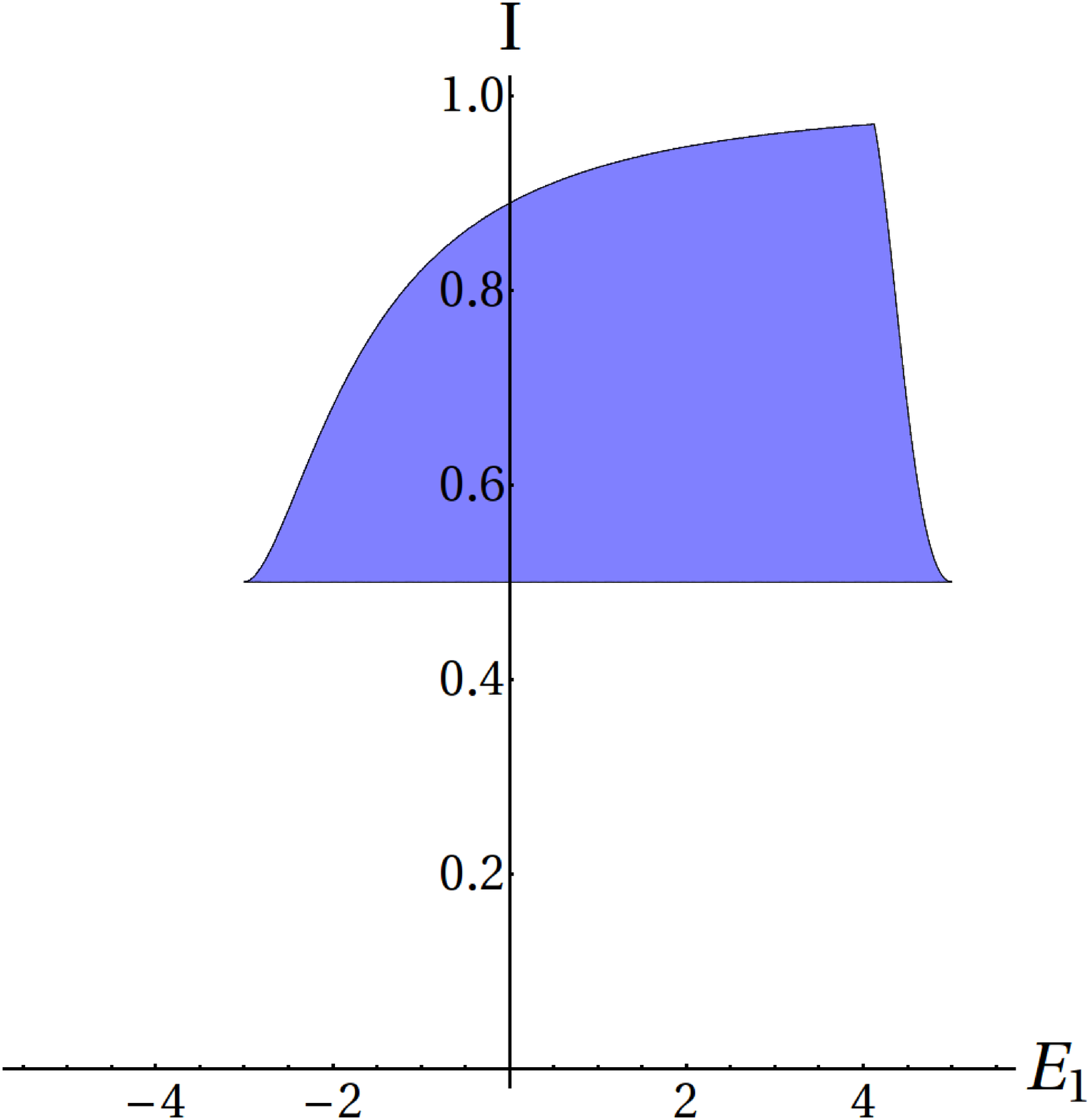} \put(0,90){(b)} \end{overpic}\\
	\end{tabular}
	\caption{The domain of (a): $p(\varepsilon_1,\varepsilon_2)$, (b): $p(E_1,I)$ for $W=8$.}

	\end{center}
\label{fig:2x2erttart}
\end{figure}

{\it Models and results -- The two-site model}
As mentioned above, an improvement to the asymptotic, large disorder limit where the one-site model works, is the so-called two-site model. We will show
here, that it gives us the main physics of Anderson model at large disorder. For such a model the Hamiltonian reads as
\begin{equation}
\cal H = \left( \begin{array}{cc}
                    \varepsilon_1 &      -1      \\
                         -1       & \varepsilon_2
                 \end{array}
         \right)\quad ,
\label{hamilton2x2}
\end{equation}
where $\varepsilon_1$ and $\varepsilon_2$ are uncorrelated random numbers drawn with uniform distribution, $p(\varepsilon_1,\varepsilon_2)=W^{-2}$ 
from the interval $[-\frac{W}{2},\frac{W}{2}]$. Consequently the support of the probability distribution is a square, that is shown in 
Fig.~\ref{fig:2x2erttart}(a).
The eigenvalues and the unnormalized eigenvectors of the Hamiltonian (\ref{hamilton2x2}) are
\begin{eqnarray} 
E_{1,2}=\frac{\varepsilon_1+\varepsilon_2}{2}\pm\sqrt{1+\left(\frac{\varepsilon_1-\varepsilon_2}{2}\right)^2}\quad ,\\
v_{1,2}=
   \left( \begin{array}{c}
            -\frac{\varepsilon_1-\varepsilon_2}{2}\mp\sqrt{1+\left(\frac{\varepsilon_1-\varepsilon_2}{2}\right)^2}\\
                                                            1
           \end{array} 
   \right)\quad .
\end{eqnarray}
Changing $\varepsilon_1$ and $\varepsilon_2$ to new variables, $t=\frac{1}{2}(\varepsilon_1+\varepsilon_2)$ and $u=\frac{1}{2}(\varepsilon_1-\varepsilon_2)$,
the probability distribution function is still constant, but on a square rotated with $45^\circ$, $p(t,u)=2/W^2$. 
Using these transformed variables the eigenenergies and the IPRs can be writen the following way
\begin{eqnarray} 
  E_{1,2}=t\pm\sqrt{1+u^2}\quad ,\\
  I=\frac{1+2u^2}{2+2u^2}\quad .
\end{eqnarray}
At this point we take the larger eigenvalue, $E_1$, and express $t$ with $E_1$ and $u$, and then express $u$ using $I$. After these two transformations 
the probability density function of $E_1$ and $I$ become
\begin{equation}
p(E_1,I) = \frac{2}{W^2}\frac{1}{\sqrt{2(1-2I)(I-1)^3}}\quad ,
\label{f2x2}
\end{equation}
whose support is quite non trivial:
\begin{equation}
\frac{1}{2} \leq I \leq 1-\frac{1}{2}\left (1+\left[\frac{\left(E_1+\frac{W}{2}\right)^2-1}{W+2E_1}\right]^2 \right )^{-1}
\label{Df2x2a}
\end{equation}
if $-\frac{W}{2}+1 \leq E_1 \leq E_0$ and
\begin{equation}
\frac{1}{2} \leq I \leq 1-\frac{1}{2}\left (1+\left[\frac{\left(E_1-\frac{W}{2}\right)^2-1}{W-2E_1}\right]^2 \right )^{-1}
\label{Df2x2b}
\end{equation}
if $E_0 \leq E_1\leq 1+\frac{W}{2}$
using $E_0$ the energy border defined in Eq.~(\ref{ener}) and appearing in Fig.~\ref{fig:1DIPR}. This domain is shown 
in Fig.~\ref{fig:2x2erttart}(b).

\begin{figure}
	\begin {center}
	\begin{tabular}{ c c }
	\begin{overpic}[width=4cm]{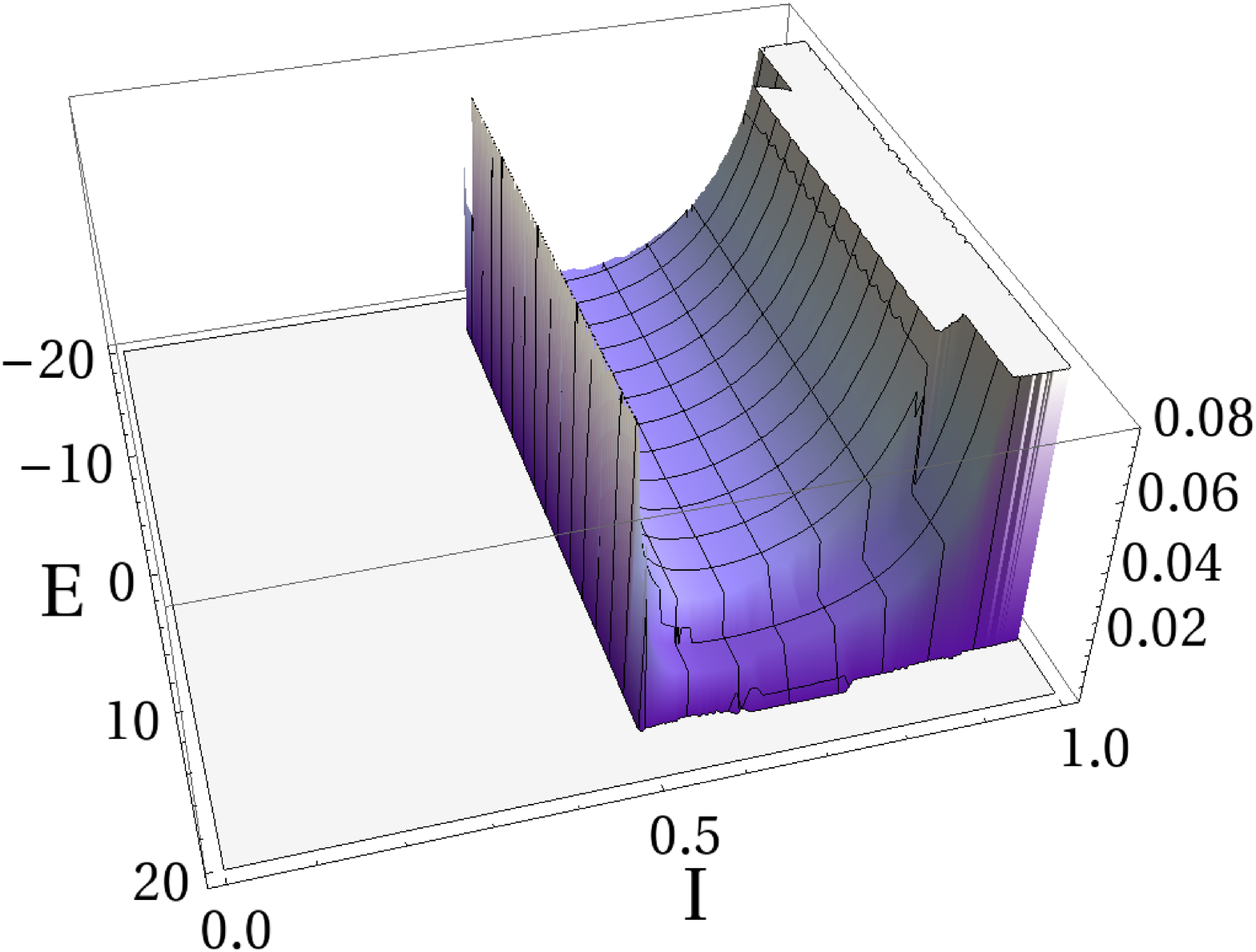} \put(-6,90){(a)} \end{overpic} & 
    \begin{overpic}[width=4cm]{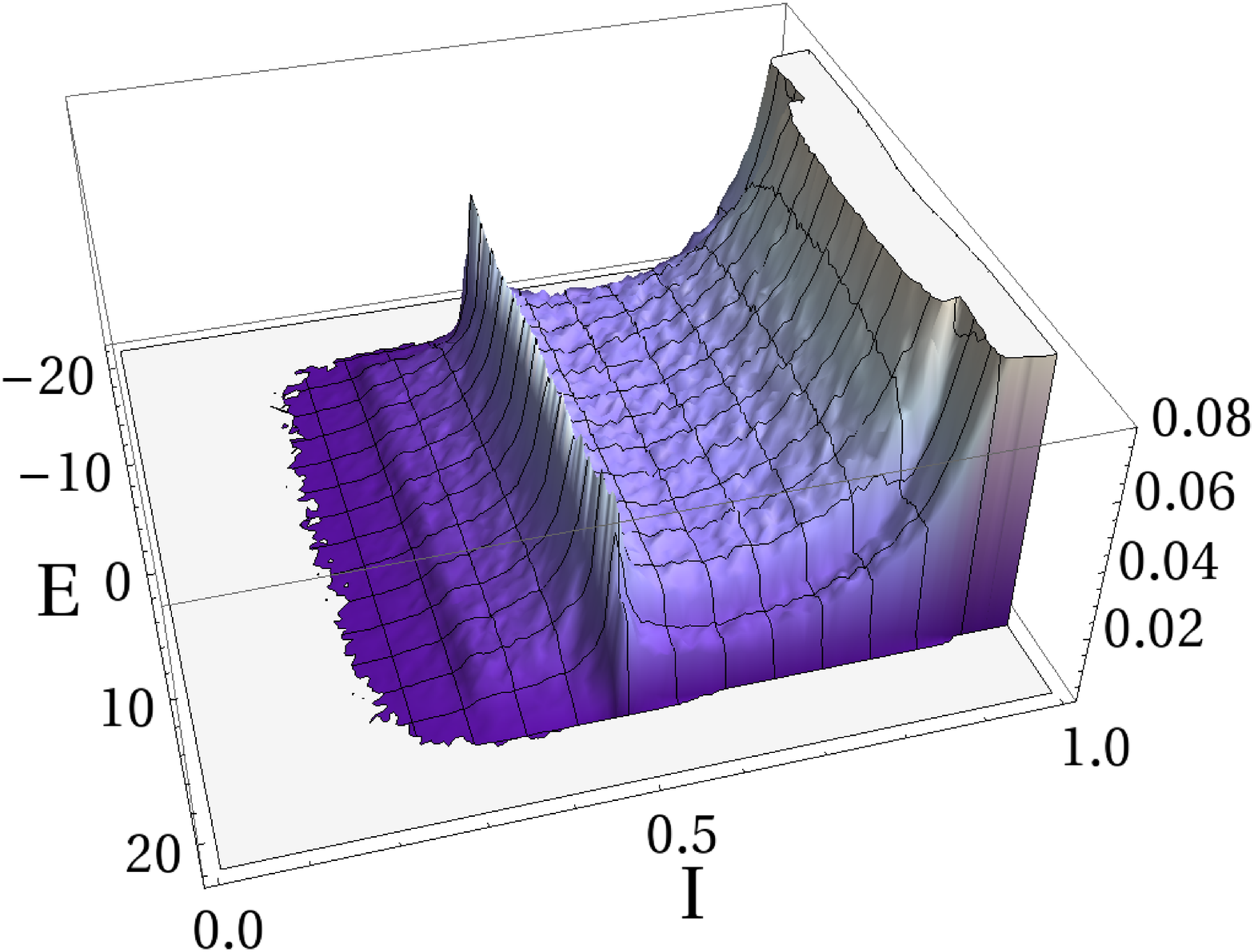} \put(-6,90){(b)} \end{overpic}\\ 
	\begin{overpic}[width=4cm]{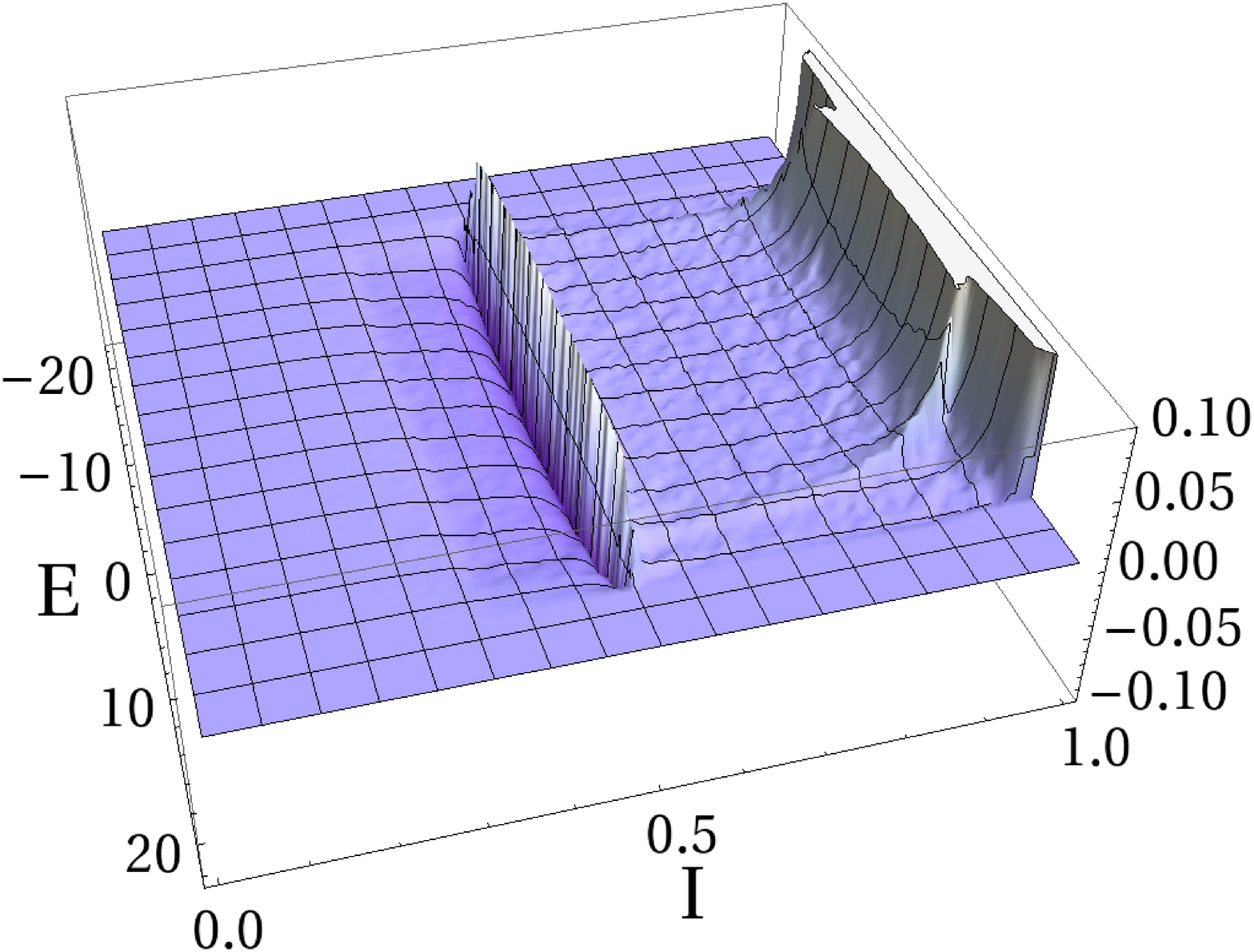} \put(-6,90){(c)} \end{overpic} & 
    \begin{overpic}[height=4cm]{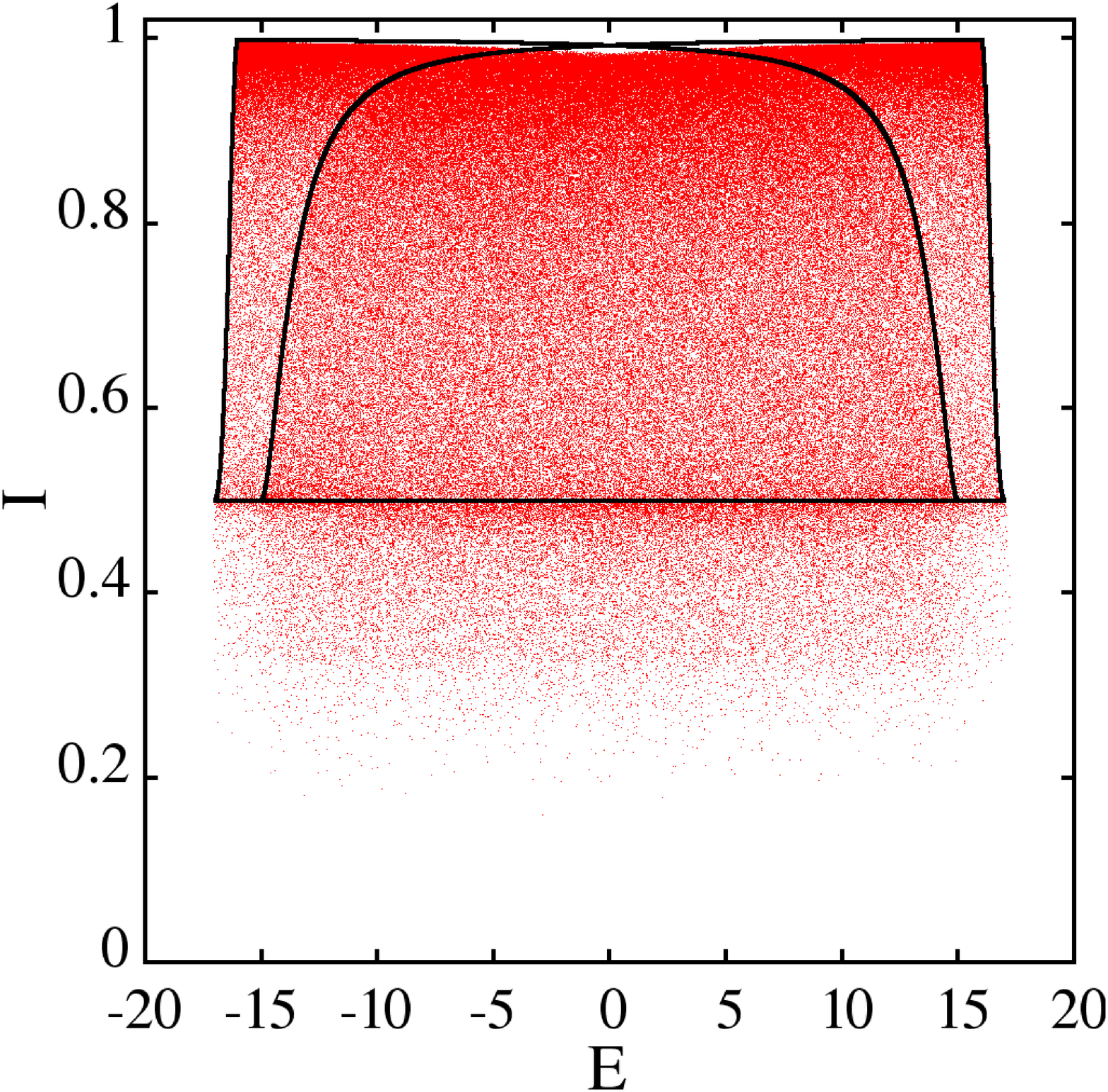} \put(-6,90){(d)} \end{overpic}
	\end{tabular}
	\caption{$p(E,I)$ (a) The analytical result of the two-site model, (b) numerical result for system size $L=512$, (c) the difference between (a) and (b).
                      (d) Red points are the numerically obtained $E$ and $I$ values at system size $L=512$, black curve is the domain of the two
                          eigenstate in the framework of the two-site model.}
	\label{fig:deIPRnumanal}
	\end{center}
\end{figure}

\begin{figure}[th]
	\begin {center}

	\begin{tabular}{ c c}
	\begin{overpic}[height=4cm]{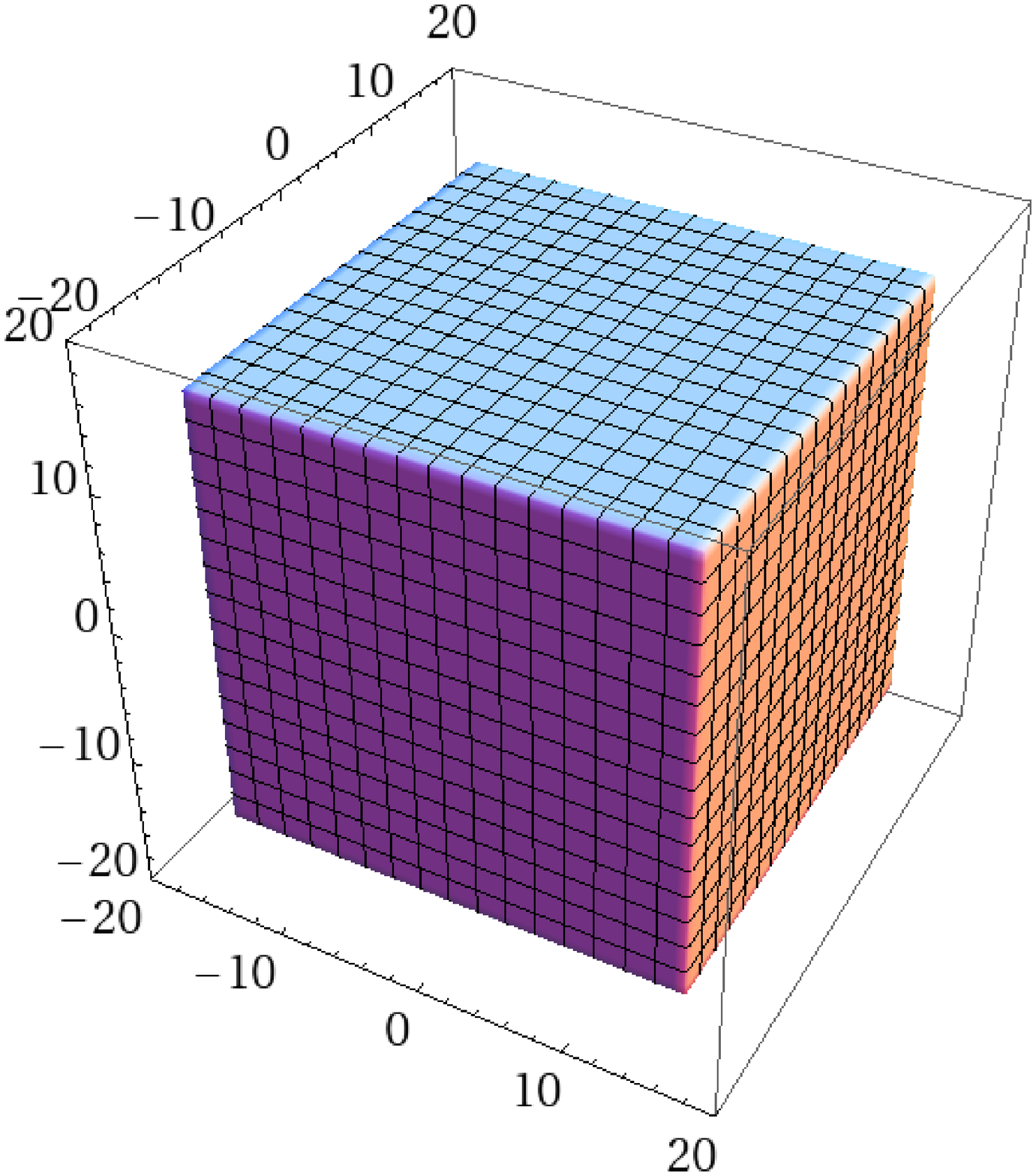}  \put(-4,90){(a)} \end{overpic} & 
    \begin{overpic}[height=4cm]{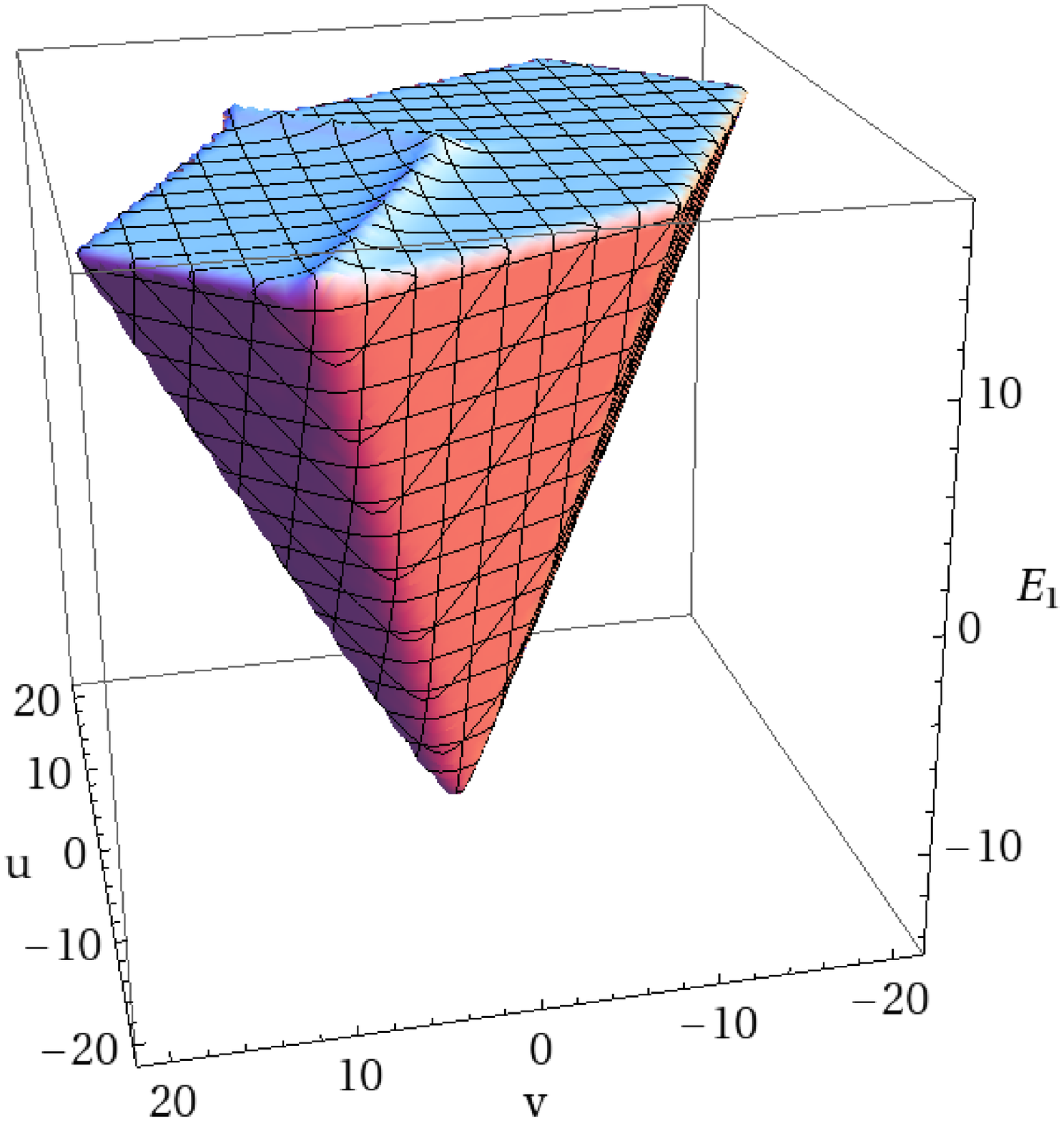} \put(-6,90){(b)} \end{overpic}\\
	\begin{overpic}[height=4cm]{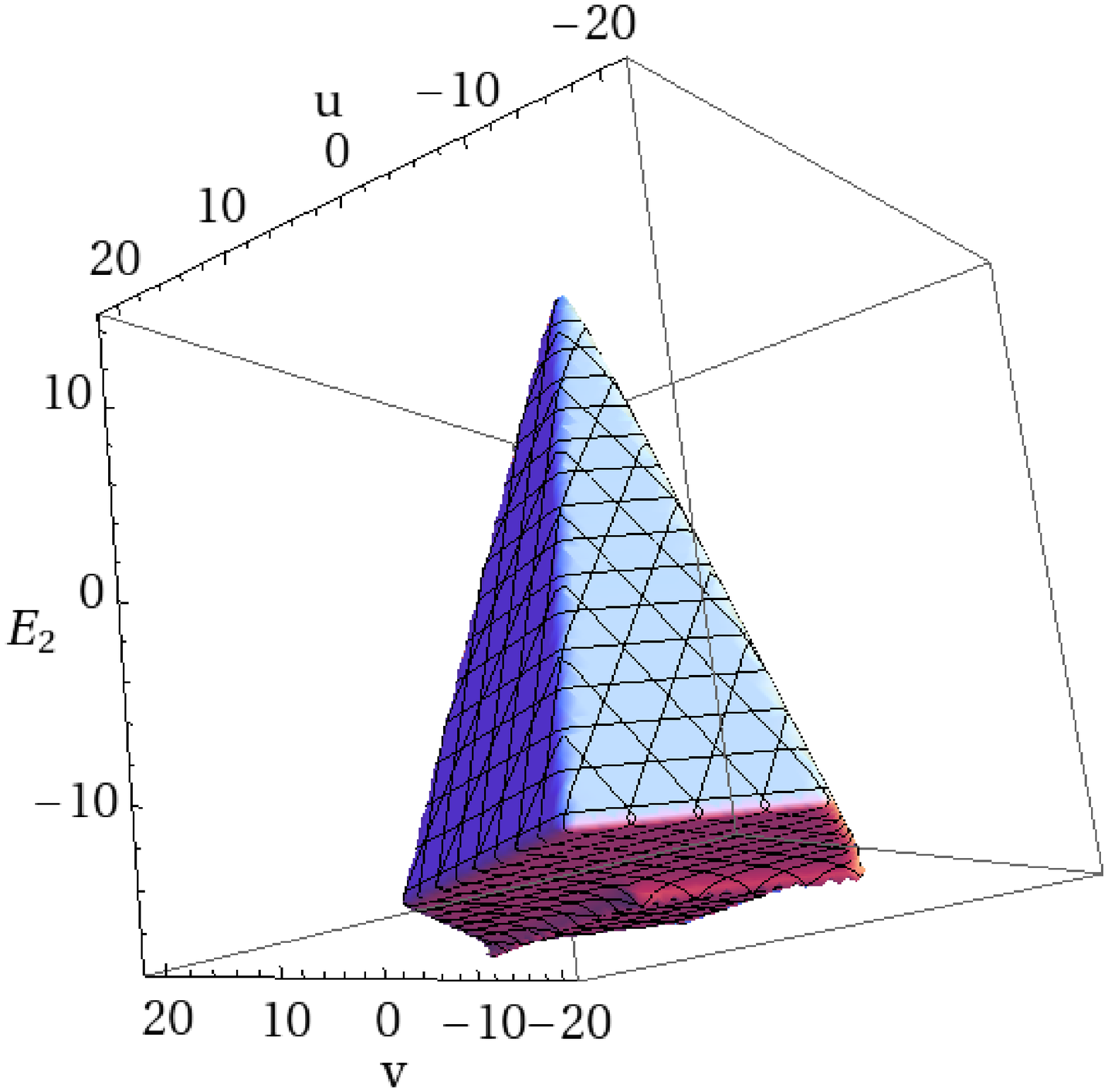} \put(-4,90){(c)} \end{overpic} & 
    \begin{overpic}[height=4cm]{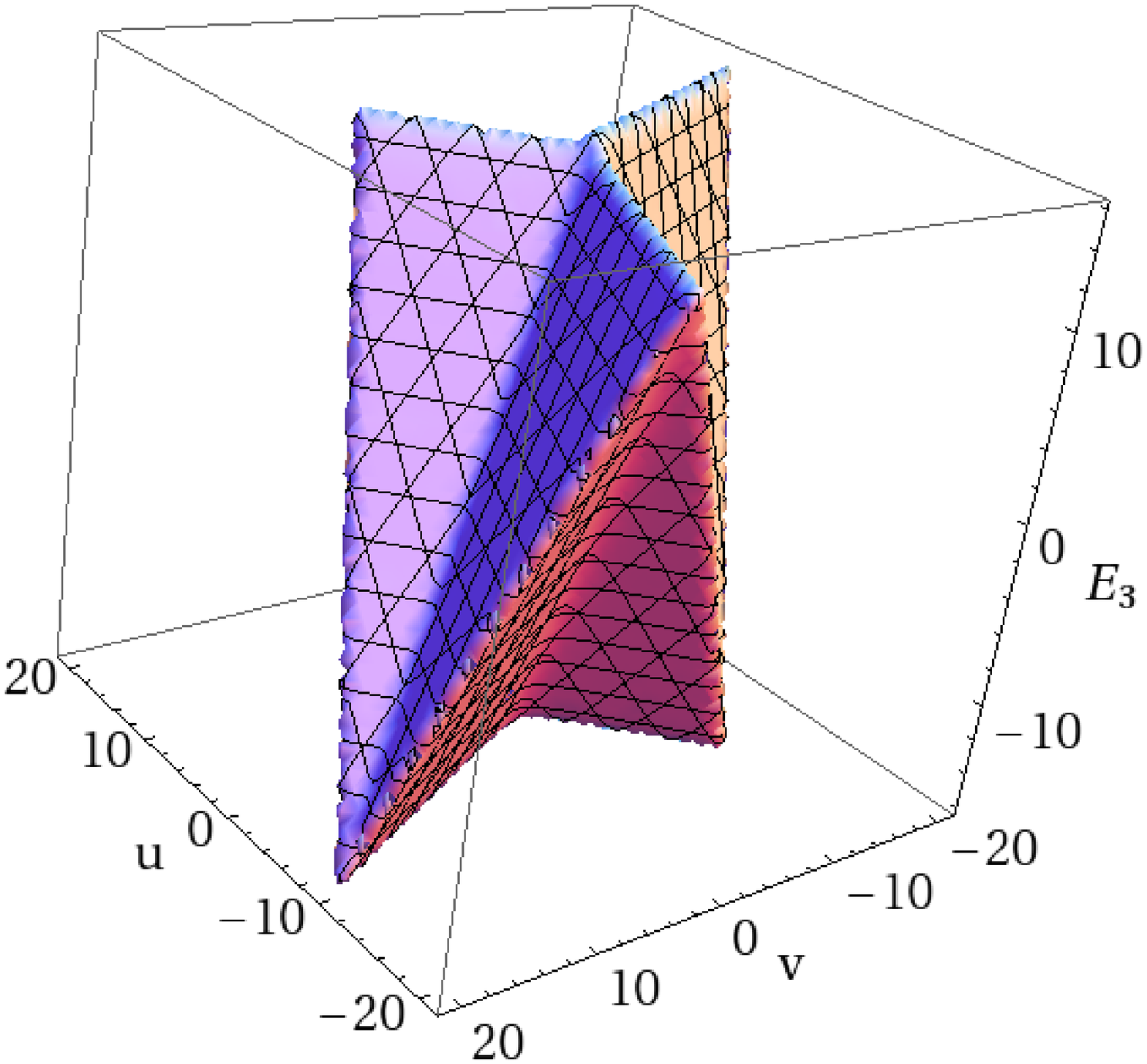} \put(-4,90){(d)} \end{overpic}\\
	\end{tabular}
	\caption{The domain of definition for variables (a) $\varepsilon_1,\varepsilon_2,\varepsilon_3$. (b) $E_1,u,v$. (c) $E_2,u,v$. (d) $E_3,u,v$.
             in a three-site model.}
	\label{fig:3x3erttart}

	\end{center}
\end{figure}

\begin{figure}[ht]
	\begin {center}

	\begin{tabular}{ c c}
	\begin{overpic}[height=4cm]{./1DIPRN512p1W32.eps} \put(-5,90){(a)} \end{overpic} & 
    \begin{overpic}[height=4cm]{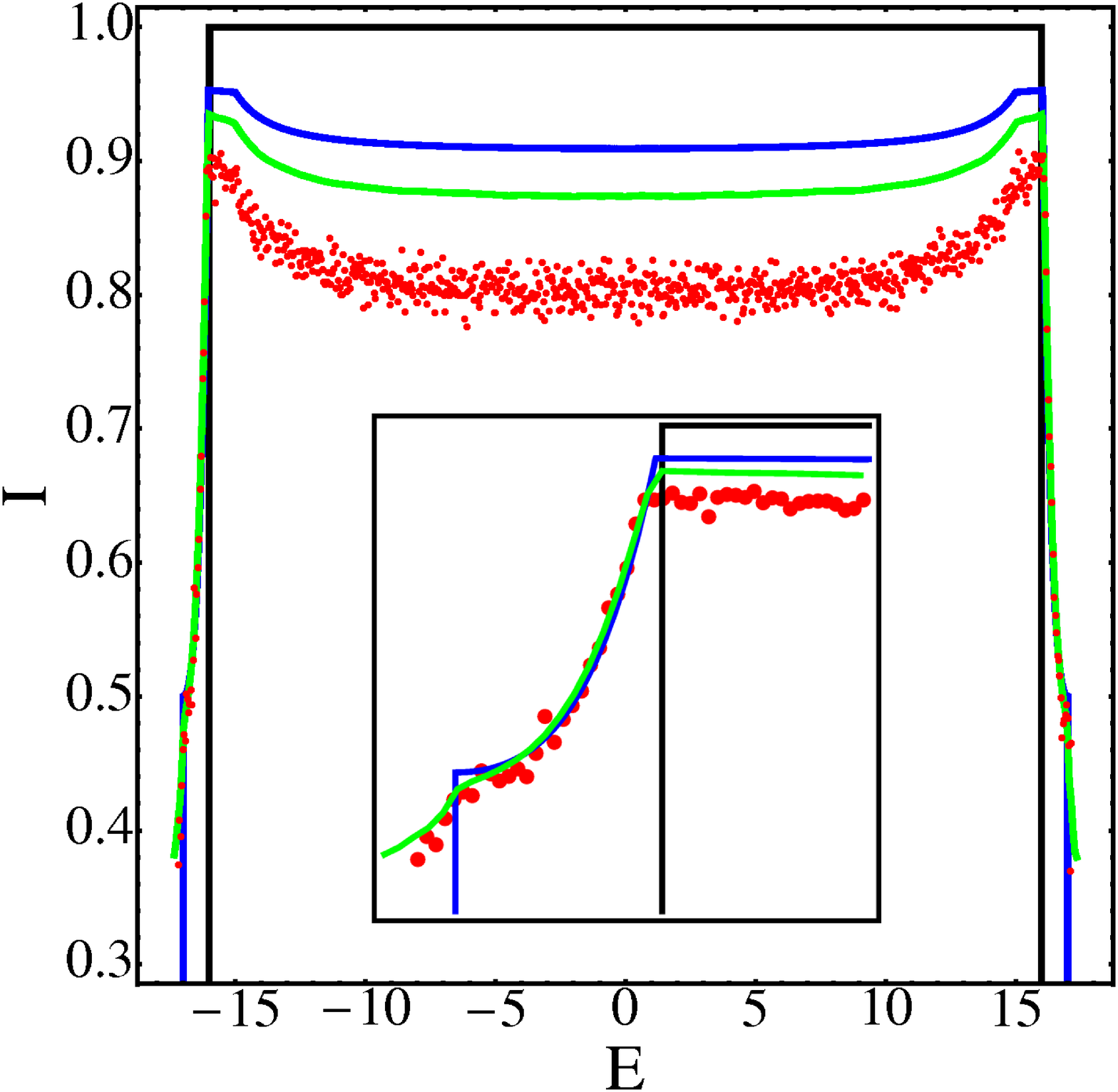} \put(-5,90){(b)} \end{overpic}\\
	\begin{overpic}[height=4cm]{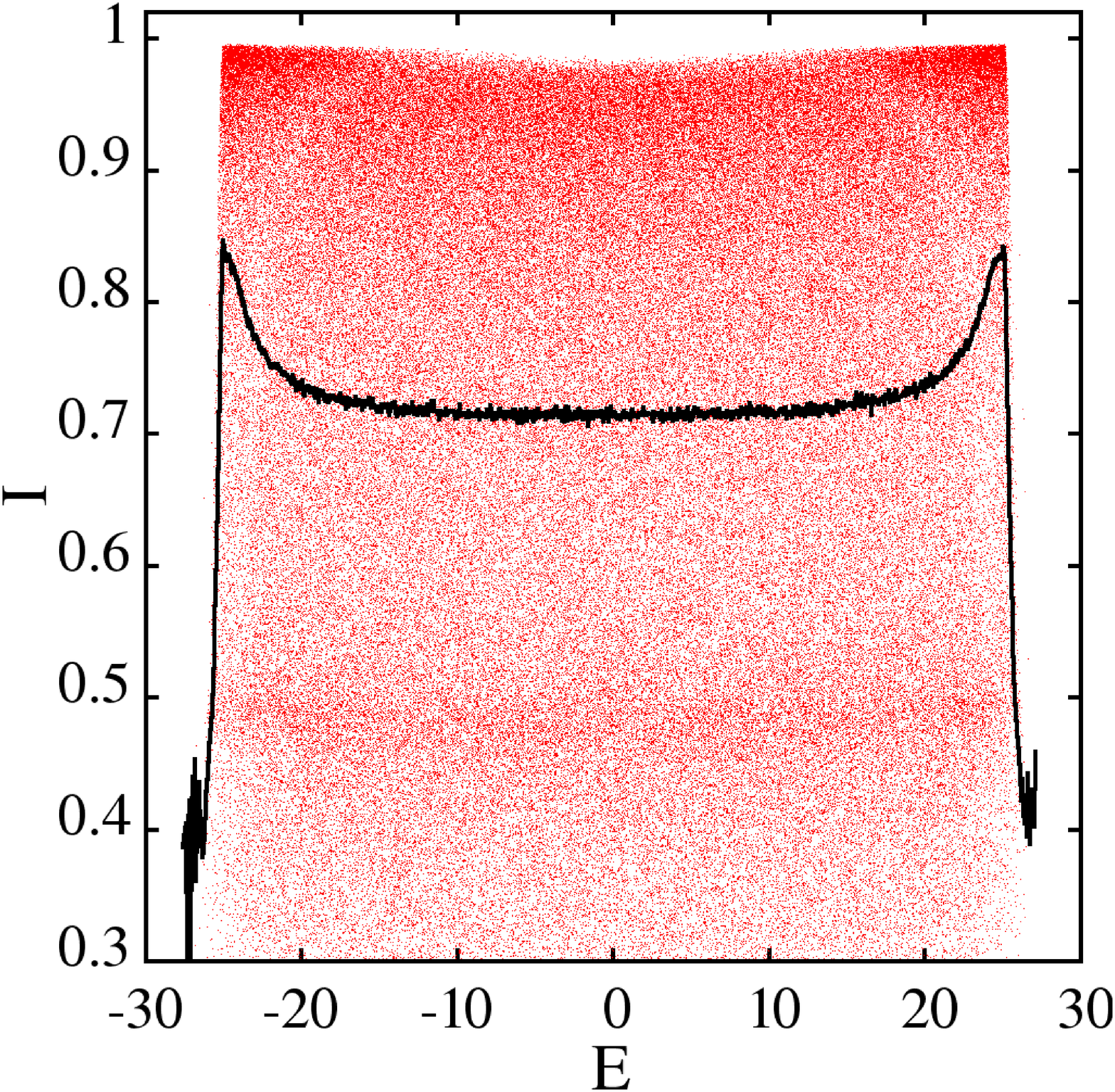} \put(-5,90){(c)} \end{overpic} & 
    \begin{overpic}[height=4cm]{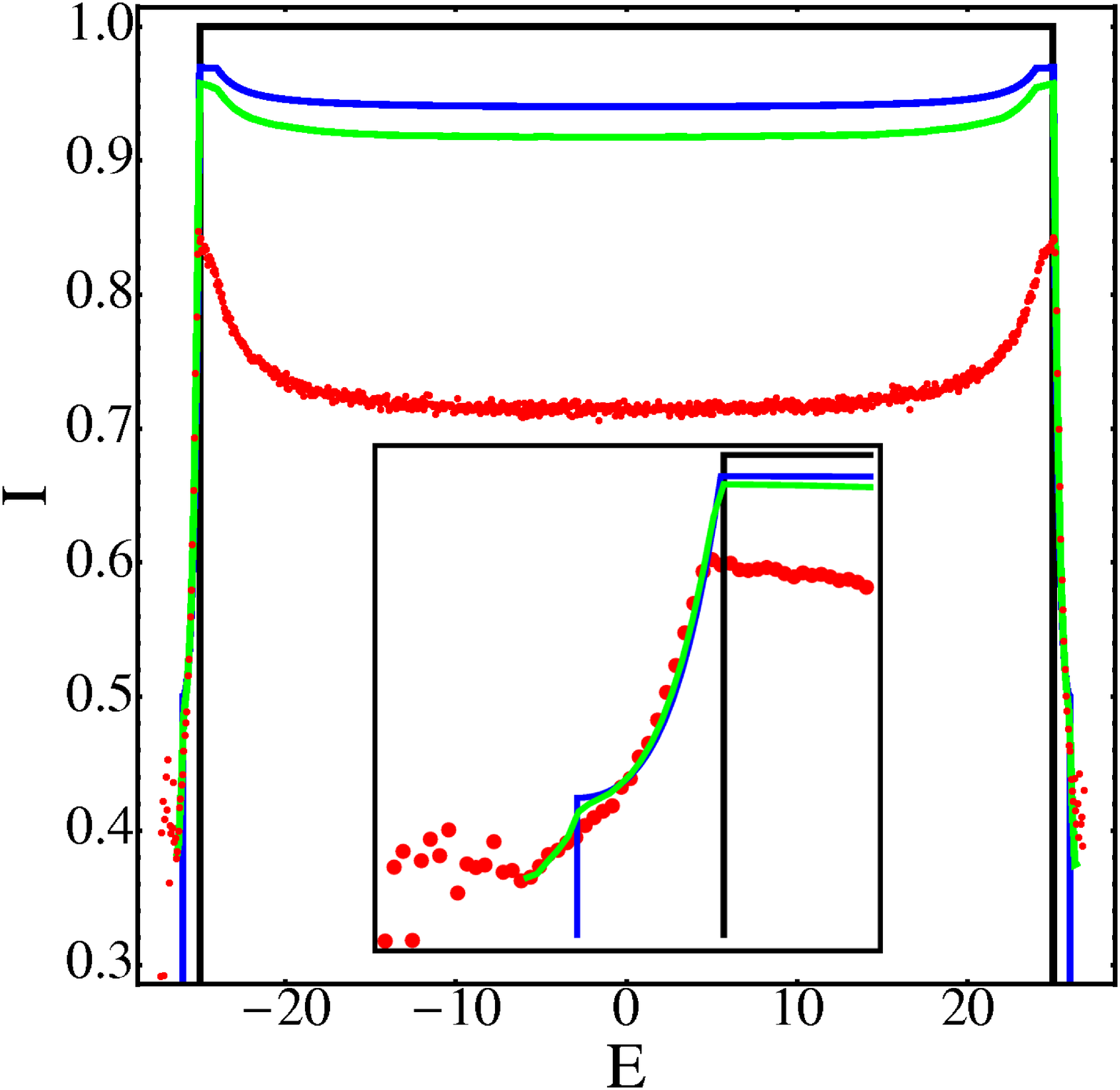} \put(-5,90){(d)} \end{overpic}\\
	\begin{overpic}[height=4cm]{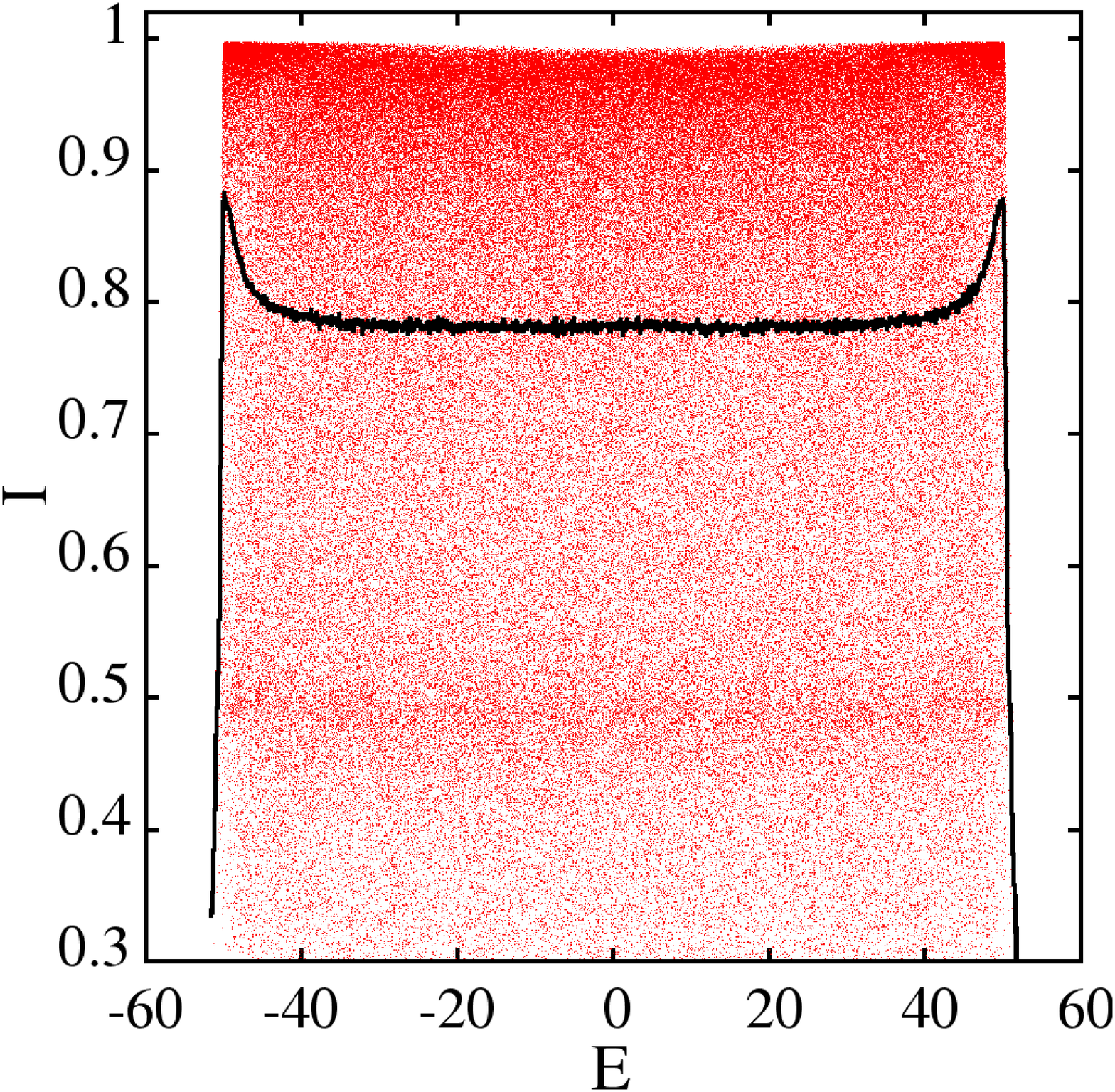} \put(-5,90){(e)} \end{overpic} & 
    \begin{overpic}[height=4cm]{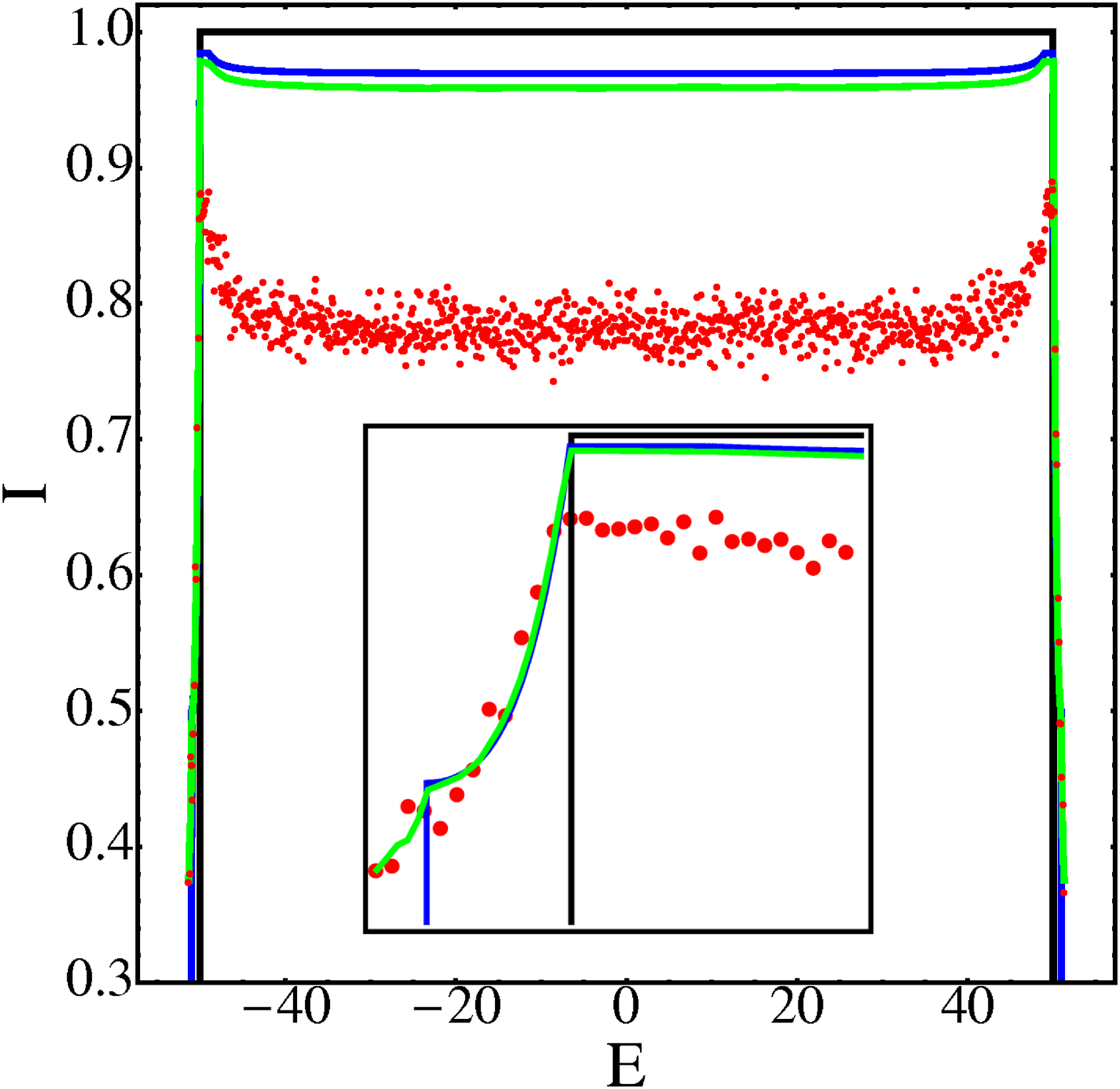} \put(-5,90){(f)} \end{overpic}\\
	\end{tabular}
	\caption{Left side: IPR as a function of energy (a) in $d=1$ with $L=512$ for $W=32$, 
                                                    (c) in $d=2$ with $L=20$ for $W=50$, 
                                                    (e) in $d=3$ with $L=8$ for $W=100$. 
              Red dots correspond to single states, black curve is the average. Right side: $\langle I\rangle$ as the function of energy 
                                                    (b) in $d=1$, (d) in $d=2$, (e) in $d=3$. 
              Dots correspond to the numerically obtained curve for a big system, black curve corresponds to the one-site model, blue to the two-site model,
              and green to the three-site model. Insets are the same, but zoomed to the left band edge.}
	\label{fig:123csucs}

	\end{center}
\end{figure}

The probability distribution function $p(E_1,I)$ is obtained for the larger eigenvalue, $E_1$. For the smaller eigenvalue, $E_2$ the result is
identical, except $E_1$ must be replaced by $-E_2$. Thus the probability density function describing the whole system is 
$p(E,I)=p(E_1,I)+p(E_2,I)=p(E_1,I)+p(-E_1,I)$, which is shown in Fig.~\ref{fig:deIPRnumanal}(a). 
If we compare this analytical function with 
Fig.~\ref{fig:deIPRnumanal}(b), the probability density function obtained numerically on a system with linear size $L=512$, the qualitative similarity 
is obvious. In Fig.~\ref{fig:1DIPR} we can clearly see, that the function is symmetric, and it is a sum of two components. In our case these two components 
are the two eigenvalues. This simple two-site model also explains the behavior of the IPR as a function of energy close to the band-edge. 
Eq.(\ref{Df2x2b}) is responsible for the decrease of the IPR depicted on the right side of Fig.~\ref{fig:2x2erttart}(b). 
Another feature of the two-site model is the peak in $p(E,I)$ at $I\approx 1/2$. However, there are some differences between the model and the 
numerical results. First of all in a two site model $I\geq 1/2$, because the state can extend maximum to two sites, but in a bigger system there exist 
a few states extending over more than two sites. Therefore in the low-$I$ regime the two-site model naturally underestimates the reality. 
Nevertheless both distributions are normalized, therefore if somewhere there is an underestimation, elsewhere there must be an overestimation, 
which gives us the hump at high $I$ values in Fig.~\ref{fig:deIPRnumanal}(c). Looking at Fig.~\ref{fig:deIPRnumanal}(d) it is clear that the two-site model 
captures very well the shape of the domain of definition for the two components, we see a little overestimation for high values of the IPR. 
From $p(E,I)$ we calculated the average of IPR, $\langle I\rangle$ as a function of the energy, $E$, which can be seen in Fig.~\ref{fig:123csucs}(b). 
The analytical curve shows a qualitative agreement with the numerical function: moving away from the band-center we see an increase in $\langle I\rangle$,
and beyond $E_0$ it decreases, showing a little shoulder. Quantitatively in the band-center the model overestimates $\langle I\rangle$, but in the decreasing
regime the two site model becomes a good approximation (see the inset of Fig.~\ref{fig:123csucs}(b). In addition the two-site model has a band of
$\left[-\frac{W}{2}-1,\frac{W}{2}+1\right]$ but obviously the real band extends beyond these limits.  {\it Models and result -- the three-site model}
The two-site model introduced in the previous subsection seems to give a qualitatively correct explanation 
for the numerically obtained distributions but as pointed out there are deficiencies. In the present 
subsection we will outline the generalization of this model to a three-site model. We will
investigate how the results change. The Hamiltonian incorporating three-sites reads as
\begin{equation} 
\cal{H} = \left( \begin{array}{ccc}
                    \varepsilon_1  &       -1      &      0       \\
                        -1         & \varepsilon_2 &     -1       \\
                         0         &       -1      & \varepsilon_3
                  \end{array} 
          \right)
\label{hamilton3x3}
\end{equation}
In the three-site model the domain of the probability density function is a cube (see Fig.~\ref{fig:3x3erttart}(a)), and the function is constant, 
$p(\varepsilon_1,\varepsilon_2,\varepsilon_3)=W^{-3}$. As in the two-site model, the eigenvectors and IPRs should not depend on the average energy, 
therefore it seems helpful to introduce new variables: $\varepsilon_1=t+u$, $\varepsilon_2=t+v$, $\varepsilon_3=t-u-v$. 
This transformation changes the domain to a parallelepiped, and the probability density function remains constant, because the transformation is linear, 
$p(t,u,v)=3/W^3$. It is easy and straightforward to compute the eigenvalues and eigenvectors of Eq.~(\ref{hamilton3x3}), but the expressions are very long, 
so we do not list here the exact expression, instead we only present their support. Every eigenvalue has the form $E_i=t+\chi_i(u,v)$, with $i=1,2,3$.
Similarly to the two-site case, the size of the eigenvectors, i.e. the IPRs depend on $u$ and $v$ only, $I(u,v)$. Picking one of the eigenvalues the 
problem can be transformed to the variables $E_i$, $u$ and $v$. The probability density function remains constant, $p(E_i,u,v)=3/W^3$. 
The difficult part of the problem is, that the domain changes to a very complicated object, which is shown in Fig.~\ref{fig:3x3erttart}(b)(c) and (d) 
for $E_1$, $E_2$ and $E_3$. 
To compute $p(E_i,I)$ we have to express $u$ as a function of $I$ and $v$, then calculate the new domain and then integrate 
over $v$. This resulted in a difficult task analytically because the expressions of the IPRs are very complicated. Instead we performed our calculation 
based on $p(E_i,u,v)=3/W^3$ using a Monte Carlo integration over the domains depicted in Fig.~\ref{fig:3x3erttart}(b),(c) and (d). 
The result is given in Fig.~\ref{fig:123csucs}(b).

In view of Fig.~\ref{fig:123csucs}(b) it is clear, that the three-site model gives a quantitatively better approximation of a large system, 
especially approaching the edge of the band (see the inset), but qualitatively the main behavior is captured already by the two-site model.

{\it Higher dimensions and summary --}
Generalization of our results to higher dimensions, $d=2,3$ and their comparisons to the numerical simulations are presented in Fig.~\ref{fig:123csucs}. 
There is a striking similarity between the figures which is due to the fact that the strongly localized regime is effectively zero dimensional, i.e. as
$W\to\infty$ the states become localized over a few sites only. The major difference is, that the line separating the two components becomes less sharp with 
increasing dimensionality. In one dimension this line can be seen very clearly, in $d=2$ it is still visible, but in $d=3$ it becomes hardly visible.

To summarize we have shown that the Anderson-model at strong localization shows interesting behavior especially approaching the band-edge. As already known the
states become more and more localized as energy increases from the band-center towards the band-edge, i.e. the inverse localization length of the states
increases as a function of energy. The IPR, on the other hand, increases up to a critical energy, $E_0$ (\ref{ener}). Beyond this limit the effective 
extension of the states can be described by a multi-site (2-site or 3-site) model because in case the eigenergy becomes larger than this critical energy,
$E>E_0$, some kinetic (hopping) energy is needed besides the random potential energy yielding in an upper bound of the IPR which in turn results in a 
decrease of the average IPR, $\langle I\rangle$ as a function of energy in this regime. In order to understand the numerical simulations we introduced
a few-site model and solved analytically capturing the main physics of the problem. 

In Ref.~\onlinecite{Johri-Bhatt} it is argued that the behavior explained in the present work is attributed to the crossover towards resonant
states similar to the effect produced by the Lifshitz-tail~\cite{Lifshitz}. It would be interesting to find the relation between our results and
the resonant states.

Financial support from by the Hungarian Research Fund (OTKA) under grants K73361 and K75529 is gratefully acknowledged.


\begin{thebibliography}{20}

\bibitem{And58}P.~W.~Anderson, Phys. Rev. {\bf 109}, 1492 (1958); A.~MacKinnon and B.~Kramer, Rep. Prog. Phys. {\bf 56}, 1469 (1993);

\bibitem{MirlinEvers}F.~Evers and A.~D.~Mirlin, Rev. Mod. Phys. {\bf 80}, 1355 (2008).

\bibitem{Johri-Bhatt}S.~Johri and R.~N.~Bhatt, arXiv:1106.1131 (2011); R.~N.~Bhatt and S.~Johri, arXiv:1204.2782 (2012).

\bibitem{VargaPH}I.~Varga, Helv. Phys. Acta, {\bf 68}, 64 (1995).

\bibitem{Lifshitz}I.~M.~Lifshitz, Adv. Phys. {\bf 13}, 483 (1964).

\end{thebibliography}
\end{document}